\documentstyle[preprint,revtex]{aps}
\begin{document}
\draft
\preprint{   }
\def\sp                                {
     \hskip 0.02in                            }
\def\ss                               {
     \scriptstyle                       }
\def\sss                               {
     \scriptscriptstyle                       }
\def\raise                             {
     {\sss +}          }
\def\k                               {
     k                            }
\def\w                               {
     \omega                             }
\def\e                               {
     \epsilon                           }
\def\a                               {
     {\bf d}                           }
\def\c                               {
     {\bf c}                           }
\def\fl {f_{\sss L}(\e)}
\def\fr {f_{\sss R}(\e)}
\def\channel                           {
     \alpha                            }
\def\ck                               {
     {\bf c}_{\k\channel}                   }
\def\ckdag                          {
     {\bf c}_{\k\channel}^{\raise}               }
\def\an {{\bf d}_n}
\def\andag {{\bf d}_n^{\raise}}
\def\am {{\bf d}_m}
\def\amdag {{\bf d}_m^{\raise}}
\def\J{ J }
\def\vkn                               {
     V_{\k\channel,n}                   }
\def\vknr                              {
     V_{\k\channel,n}^R                   }
\def\vn                              {
     V_{\channel,n}(\e)                   }
\def\vm                              {
     V_{\channel,m}(\e)                   }
\def\vmstar                              {
     V_{\channel,m}^*(\e)                   }
\def\vml                              {
     V_{\channel,m}(\e)                   }
\def\vkm                               {
     V_{\k\channel,m}                   }
\def\vkmstar                               {
     V_{\k\channel,m}^*                   }
\def\gaml {
 \Gamma_{\channel;m,n}^{\sss L}(\e)            }
\def\Gamr {
 {{\bf \Gamma}^{\sss R}} }
\def\Gaml {
 {{\bf \Gamma}^{\sss L}}    }
\def\Gam {
 {\bf \Gamma}    }
\def\gamr {
 \Gamma_{\channel;m,n}^{\sss R}(\e)            }
\def\Gr {
 {\bf G}^r }
\def\Ga {
 {\bf G}^a }
\def\Gl {
 {\bf G}^< }
\def\amn {
 A_{mn}(\e)       }
\def\spin                           {
     \sigma                            }
\def\aspin                               {
     {\bf d}_{\spin}                   }
\def\aspindag                          {
     {\bf d}_{\spin}^{\raise}               }
\def\ckspin                               {
     {\bf c}_{\k\spin}                   }
\def\ckspindag                            {
     {\bf c}_{\k\spin}^{\raise}              }
\def\espin                           {
     \epsilon_{\spin}                   }
\def\ezero                           {
     \e_{\sss 0}                }
\def\kin                            {
     \k,\channel{\sss \in L,R}                      }
\def\channelin                            {
     \channel{\sss \in L,R}                      }
\def\ndown                            {
     n_{\sss\downarrow}                         }
\def\nup                            {
     n_{\sss\uparrow}                         }
\def\vkspin {V_{\k\spin} }
\def\cond                             {
     G                               }
\def\final { i}
\def\efinal{ E_{\final}}
\def\Gnkl{ G_{n,\k\channel}^{\, <} (\w)}
\def\Gknl{ G_{\k\channel,n}^{\, <} (\w)}
\def\Gmnl{ G_{m,n}^{\, <} (\w)}
\def\Gnml{ G_{n,m}^{\, <} (\w)}
\def\Gnmg{ G_{n,m}^{\, >} (\w)}
\def\Gnmle{ G_{n,m}^{\, <} (\e)}
\def\Gnmad{ G_{n,m}^{\, a} (\e)}
\def\Gnmre{ G_{n,m}^{\, r} (\e)}
\def\Gnmge{ G_{n,m}^{\, >} (\e)}
\def\Gnmt{ G_{n,m}^{\, t} (\w)}
\def\Gnmt{ G_{n,m}^{\, t} (\w)}
\def\Gmntt{ G_{m,n}^{\, \tilde{t}} (\w)}
\def\gkkl{ g_{\k\channel,\k\channel}^{\, <} (\w)}
\def\gkkg{ g_{\k\channel,\k\channel}^{\, >} (\w)}
\def\gkkt{ g_{\k\channel,\k\channel}^{\, t} (\w)}
\def\gkktt{ g_{\k\channel,\k\channel}^{\, \tilde{t}} (\w)}
\def\F{ {\bf F}}
\def\lefta{<}
\def\righta{>}
\def\DeltaL                             {
    \Delta_{\ss L}                              }
\def\DeltaR                             {
    \Delta_{\ss R}                              }
\def\DeltaLR                             {
    \Delta_{\ss L/R}                              }
\def\ALR                             {
    A_{\ss L/R}                              }
\def\JL                                 {
    J_{\ss L}                              }
\def\JR                                 {
    J_{\ss R}                              }
\def\Jdc                                {
    J_{\rm dc}                              }
\def\mul                                 {
    \mu_{\ss L}                              }
\def\mur                                 {
    \mu_{\ss R}                              }
\def\mulr                                 {
    \mu_{\ss L/R}                              }
\def\lr{1}
\def\alw {2}
\def\dbrts {3}
\def\set {4}
\def\guimar{5}
\def\fu{6}
\def\sokol{7}
\def\mw{8}
\def\keldysh {9}
\def\dysoneq {10}
\def\wjw     {11}
\def\accum   {12}
\def\tlogt   {13}
\def\chena {14}
\def\leoa    {15}
\def\squote{}
\def\quote#1#2#3#4{\squote {#1,\ {\sl#2}\/ {\bf#3}, #4}.\par}
\def\qquote#1#2#3#4{\squote {#1,\ {\sl#2}\/ {\bf#3}, #4};}
\def\nquote#1#2#3#4{\squote {#1,\ {\sl#2}\/ {\bf#3}, #4}}
\def\book#1#2#3{\squote { #1,\ in {\sl#2}, edited by #3}.\par}
\def\nbook#1#2#3{\squote { #1,\ in {\sl#2}, edited by #3}}
\def\bbook#1#2#3{\squote { #1,\ in {\sl#2}, edited by #3}.}
\def\trans#1#2#3{\left[ {\sl #1} {\bf #2},\ #3\right]}
\def\apl{Appl. Phys. Lett.}
\def\prl{Phys. Rev. Lett.}
\def\prb{Phys. Rev. B}
\def\pr{Phys. Rev.}
\def\rPoint {}
\def\lPoint                        {}
\def\lPoint {}
\def\cl{\centerline}
\def\dl{\displaylines}

\newcount\rnumber
\def\defqoute#1{\newcount#1 #1=\rnumber\global\advance\rnumber by 1}
\rnumber=1
\def\rpoint                          {     $====>$                        }
\def\lpoint                          {     $<====$                        }
\begin{title}
Time-Dependent Transport through a Mesoscopic Structure
\end{title}
\author{Ned S. Wingreen}
\begin{instit}
NEC Research Institute, 4 Independence Way, Princeton, NJ 08540
\end{instit}
\author{Antti-Pekka Jauho$^*$}
\begin{instit}
NORDITA, Blegdamsvej 17, DK-2100 Copenhagen, Denmark
\end{instit}
\author{Yigal Meir}
\begin{instit}
Department of Physics, University of California, Santa Barbara, CA 93106
\end{instit}
\begin{abstract}
We present a general formulation of the nonlinear, time-dependent current
through a small interacting region, where electron energies are
changed by time-dependent voltages.
An exact solution is obtained for the
non-interacting case when the elastic coupling to the leads
is independent of energy. Temporal phase coherence
in a double-barrier tunneling structure produces ``ringing" in
the response of the current to a voltage pulse, which can be
observed experimentally
in the dc-current by varying the pulse length in a train
of voltage pulses. The nonlinear current due to an ac-bias also
shows complex time-dependence.
\end{abstract}
\pacs{PACS numbers: 73.20.Dx 73.40.Ei 73.40.Gk 73.50.Fq}

The importance of the spatial coherence of electronic wavefunctions
is one of the hallmarks
of the mesoscopic regime. A panoply of mesoscale phenomena including
weak-localization and weak anti-localization,$^{\lr}$ Aharonov-Bohm
oscillations,$^{\alw}$ and Universal conductance fluctuations$^{\alw}$
all rely on
the phase coherence of electrons in small structures. Since the
traditional experimental probe of these effects is steady-state transport,
the role of {\it temporal} phase coherence is generally subsumed under the
effects of spatial coherence. However, recent experimental
progress$^{\dbrts-\guimar}$
has opened the door to direct measurement of
electronic phase coherence in time.

Interest in the time domain has sprung both from device
applications of double-barrier resonant-tunneling structures,$^{\dbrts}$ and
from possible current standard applications of single-electron tunneling
circuits.$^{\set}$ While the interactions among electrons are mean-field
like in the three-dimensional resonant-tunneling structures, the
low-dimensional confinement of electrons in the single-electron
circuits results in strong many-body correlations. A general theory
for temporal coherence in these structures must therefore rigorously
account for interactions.

In this letter, we first present a general formulation of the nonlinear,
time-dependent current through a small interacting region coupled to two
non-interacting leads.$^{\fu}$ A general result,
Eq. (\ref{eq:JLb}), expresses the time-dependent current
flowing into the interacting
region from one lead in terms of local Green functions.
This expression is then used to explore
temporal coherence in the response of a mesoscopic system to
time-dependent external
driving. We focus on a fully nonlinear, exactly solvable example ---
non-interacting electrons traversing a double-barrier tunneling structure
with energy independent coupling to the leads.
The temporal coherence of this system is evident in the ``ringing"
of the current in response to a rectangular pulse of the bias.
This ringing
can be observed experimentally in the {\it dc-current}
by varying the pulse length in a train of voltage pulses applied
to the structure. Similarly, the current flowing in response to an ac-bias
displays complex time-dependence, which is reflected
in oscillations of the dc-current vs.~{\nobreak driving}~frequency.$^{\sokol}$

The formalism and Hamiltonian are similar to those employed for
the steady-state current in Ref. {\mw},
\begin{equation}
H = \sum_{k,\alpha\in L/R} \!\!\!\epsilon_{k\alpha}(t)
{\bf c}_{k\alpha}^{\dagger} {\bf c}_{k\alpha}
\,+\, H_{\rm int}[\lbrace{\bf d}_n\rbrace,\lbrace{\bf d}_n^{\dagger}\rbrace,t]
\,\,+ \!\!\sum_{k,\alpha\in L/R\atop n}
[V_{k\alpha,n}(t){\bf c}_{k\alpha}^{\dagger} {\bf d}_n + h.c. ],
\label{eq:H}
\end{equation}
where $\ckdag (\ck)$ creates (destroys) an electron with momentum
$k$ in channel $\channel$ in either the left ($L$) or the right ($R$) lead,
and $\{\andag\}$ and $\{\an\}$ form a complete, orthonormal set
of single-electron creation and annihilation operators in the
interacting region. However, time dependence due to external
driving is now explicitly included in the energies of states
in the interacting region and in the leads,
in the hopping matrix elements, and
in the interactions themselves.  A central assumption
is that interactions between electrons in the leads
and between electrons in the leads and in the interacting region
can be neglected. Geometrically, the leads must therefore rapidly
broaden into large metallic contacts in which interactions are
strongly screened.
Since transport experiments on mesoscale structures
typically satisfy this requirement, the model is directly relevant
to experiment.
Note that because time-dependent potentials
must be associated with changes of charge, there will be capacitive
currents flowing in the contacts in addition to the currents
from the interacting region determined by (\ref{eq:H}).
The contribution of the capacitive currents can be eliminated,
however, by measuring only the time-averaged current.

The Keldysh approach$^{\keldysh}$ to calculating the current
flowing into and out of the interacting region
treats the contacts as systems separately in equilibrium in the
distant past, possibly
with different chemical potentials.
Physically, applying a time-dependent bias (electrostatic-potential
difference) changes the energies of states in
the leads via $\epsilon_{k\alpha}(t)$,
without changing their occupations.  This preserves the
coherent evolution of phase in the leads, $\Psi_{k\alpha}(t) \propto
\exp[-i\int dt'\,\epsilon_{k\alpha}(t')]$,
and produces interference in tunneling between the leads and the
interacting region, due to their different time-dependent energies.

The time-dependent current from the
left lead into the interacting region is
\begin{equation}
J_L(t)= {2e\over\hbar}\, {\rm Re}\, \Bigl\lbrace\!\sum_{k,\alpha\in L\atop n}
V_{k\alpha,n}(t)
G_{n,k\alpha}^<(t,t)\Bigr\rbrace,
\label{eq:JLa}
\end{equation}
where in (\ref{eq:JLa})
we have just rewritten the expectation of the current matrix element
in terms of the Keldysh Green function$^{\keldysh}$
$G_{n,k\channel}^<(t,t') \equiv i\langle\ckdag(t') \an(t) \rangle$.
Since the Hamiltonian describing the leads is non-interacting, one
has the Dyson equation
\begin{equation}
G_{n,k\alpha}^<(t,t')
=
\sum_m \int dt_1
V_{k\alpha,m}^*(t_1)
[G^r_{n,m}(t,t_1)g^<_{k\alpha,k\alpha}(t_1,t')
+G^<_{n,m}(t,t_1)g^a_{k\alpha,k\alpha}(t_1,t') ],
\label{eq:Glessa}
\end{equation}
where $G_{n,m}^<(t,t') \equiv i\langle\amdag(t') \an(t) \rangle$,
and $G_{n,m}^r(t,t') \equiv
-i\theta(t-t')\langle\{\amdag(t'), \an(t)\} \rangle$
is the retarded Green function.
The time-dependent
Green functions in the leads for the uncoupled system, which
appear in (\ref{eq:Glessa}), are given by
\begin{eqnarray}
&g_{k\alpha,k\alpha}^<(t,t') &= i f(\epsilon^0_{k\alpha})
\exp\Bigl[ -i\! \int_{t'}^t dt_1\epsilon_{k\alpha}(t_1)\Bigr]
\cr
&g_{k\alpha,k\alpha}^a(t,t') &= i\theta( t' - t)\exp\Bigl[
 -i\! \int_{t'}^t dt_1\epsilon_{k\alpha}(t_1)\Bigr],
\label{eq:gleads}
\end{eqnarray}
where $\epsilon_{k\alpha}^0$ is the energy of the state $k$,
with occupation
$f(\epsilon^0_{k\alpha})$,
when the system was prepared in the distant past,
and the advanced Green function is related to the
retarded Green function by
$g^a(t,t')  =
[g^r(t',t)]^*$.
It is convenient in (\ref{eq:JLa}) to turn
the sum over momentum states $k$ in the
leads into an integral over energies, and to define the elastic
coupling between the leads and the states in the
interacting~{\nobreak region}~via
\begin{equation}
\big [ \Gamma^L(\epsilon,t',t) \big ]_{m,n}=
2\pi \sum_{\alpha\in L} \rho_{\alpha}(\epsilon)
V_{\alpha,n}(\epsilon,t)V^*_{\alpha,m}(\epsilon,t')
\exp\Bigl[\,i\!\int_{t'}^tdt_1\Delta_{\alpha}(\epsilon,t_1)\Bigr],
\label{eq:GammaL}
\end{equation}
where $\rho_{\alpha}(\epsilon)$ is the density of states in channel
$\channel$ and the energy, $\e_{k\alpha}(t)$, of each state in the leads is
separated into a constant part
$\epsilon_{k\alpha}^0 = \e$
and a time-dependent part
$\Delta_{\channel}(\epsilon,t)$.
Rewriting the current in terms of the elastic coupling, and using
matrix notation for the level indices in the interacting region,
we find
\begin{equation}
J_L(t)=-{2e\over\hbar}
\int_{-\infty}^t \!\!dt' \int {d\epsilon\over2\pi}
\,{\rm ImTr} \lbrace
{\rm e}^{i\epsilon(t-t')}
{\bf \Gamma}^L(\epsilon,t',t)
[{\bf G}^<(t,t')+f_L(\epsilon){\bf G}^r(t,t')]\rbrace.
\label{eq:JLb}
\end{equation}
An analogous expression applies for the current flowing
in from the right lead.

Eq. (\ref{eq:JLb}) is the central formal result of this work.
The strong resemblance to the steady-state result in Ref. (\mw)
means that the time-dependent problem is not significantly harder
than the time-independent one. For a range of mesoscopic systems,
a rigorous formula for the steady-state current is a
useful tool, and we believe (\ref{eq:JLb}) will be a useful
tool to study dynamics.  While
interacting quantum transport is addressable with (\ref{eq:JLb}),
we use it here to explore dynamics
in an exactly solvable, non-interacting system
corresponding to a quantum-well structure.

In general, if interactions are neglected,
the retarded and advanced Green functions are given by
standard Dyson equations,$^{\dysoneq}$ and the
Keldysh Green function in (\ref{eq:JLb})
is related to them via
\begin{equation}
{\bf G}^<(t,t')  = i\int\! dt_1\int\! dt_2\, {\bf G}^r(t,t_1)
\Bigl[\sum_{L,R}\int {d\epsilon\over 2\pi}
e^{i\epsilon (t_2-t_1)}f_{L/R}(\epsilon)
{\bf \Gamma}^{L/R}(\epsilon,t_1,t_2)\Bigr]
{\bf G}^a(t_2,t').
\label{eq:Glessb}
\end{equation}
A tractable, non-interacting example of particular interest corresponds to a
double-barrier tunneling structure containing a single
resonant level, $\epsilon_0(t) = \epsilon_0 + \Delta(t)$,
with energy-independent coupling to two leads,
\begin{equation}
\Gamma_{L/R}(t',t)=\Gamma_{L/R}\,
\exp\Bigl[\,i\!\int_{t'}^t dt_1 \Delta_{L/R}(t_1)\Bigr].
\label{eq:GammaLR}
\end{equation}
In (\ref{eq:GammaLR}), the time dependence in the leads is
restricted to a rigid shift of all states in the left(right) lead by
$\Delta_{\ss L(R)}(t)$.
Within this framework, application of
a voltage bias corresponds to a shift of energies in one lead
with respect to the other, and generally a shift of the resonant
level as well.
For energy-independent coupling to the leads, the retarded
Green function for the resonant level is
independent of energy shifts $\Delta_{L/R}(t)$ in the leads,
and is given by$^{\wjw}$
$G^r(t,t')=
\exp[-\Gamma(t-t')/2]\, g^r(t,t')$,
where $g^r(t,t')$ is the retarded Green function for the uncoupled level,
\begin{equation}
g^r(t,t') = -i\theta(t - t')\,
\exp\Bigl[-i\!\int_{t'}^tdt_1\epsilon_0(t_1)\Bigr],
\label{eq:Grb}
\end{equation}
and where $\Gamma\equiv\Gamma_L+\Gamma_R$ is the total elastic
coupling to the leads.

The resulting expression
for the retarded
Green function of the resonant level,
$G^r(t,t')$,
can be used in (\ref{eq:Glessb})
to generate the Keldysh Green function, $G^<(t,t')$, which contains
all the information concerning non-equilibrium occupations.
Physical quantities such as the occupancy of the
level and the currents through the barriers
can then be expressed directly in terms of $G^<(t,t')$,
or, more conveniently via (\ref{eq:Glessb}),
in terms of $G^r(t,t')$ and the occupation
functions in the leads. The occupancy of the level,
$n(t) = -i G^<(t,t)$, and the current, $J_{L/R}$ from (\ref{eq:JLb}), are
given by:
\begin{equation}
n(t)  = \sum_{L/R}\Gamma_{L/R}
\int {d\epsilon \over 2 \pi} f_{L/R}(\epsilon)
|A_{L/R}(\epsilon,t)|^2
\label{eq:N}
\end{equation}
\FL
\begin{equation}
J_{L/R}(t)  = -{e\over\hbar} \Gamma_{L/R}
\Bigl[\,n(t) +  \int {d\epsilon\over\pi}
f_{L/R}(\epsilon) {\rm Im}\lbrace A_{L/R}(\epsilon,t)\rbrace \Bigr],
\label{eq:JLR}
\end{equation}
where we have defined
\FL
\begin{equation}
A_{L/R}(\epsilon,t) =
\int_{-\infty}^t dt_1
\exp\Bigl[i\!\int^t_{t_1}dt_2(\epsilon+\Delta_{L/R}(t_2))\Bigr]
 G^r(t,t_1).
\label{eq:ALR}
\end{equation}
In the time-independent case,
$\ALR(\epsilon,t)$ is the Fourier transform of
the retarded Green function, and
one finds the usual result for the resonant-tunneling current,
\begin{equation}
J = {e\over h} \int\! d\epsilon\, [f_L(\epsilon)-f_R(\epsilon)]
{\Gamma_L \Gamma_R \over (\epsilon-\epsilon_0)^2+(\Gamma/2)^2}.
\label{eq:Jusual}
\end{equation}

While arbitrary time dependence of both the bias and level energy can
be addressed via Eqs. (\ref{eq:Grb}-\ref{eq:ALR}), we have
chosen to consider the responses to a rectangular pulse
and to an ac-bias
as examples of experimental relevance. As discussed above,
an important experimental caveat is that the time-dependent
current flowing in the contacts will include capacitive contributions from
the accumulation and depletion layers on either side of
the tunneling barriers.$^{\accum}$
It is possible, however, to measure exclusively the
current flowing through the barriers by measuring
the time-averaged current, since the dc-current must
be uniform throughout the structure.

{}From Eq. (\ref{eq:ALR}), one finds for
a rectangular pulse of duration $s$ starting at $t=0$,
\begin{equation}
A_{L/R}(\epsilon,t) =
{ {\epsilon - \epsilon_0 + i\Gamma/2
- (\Delta - \Delta_{L/R})
 \Bigl[1-e^{i(\epsilon-\epsilon_0+i\Gamma/2)(t-s)}
\Bigl(1-e^{i(\epsilon-\epsilon_0-\Delta+\Delta_{L/R}+i\Gamma/2)s}
\Bigr)\Bigr]   }
 \over
{(\epsilon - \epsilon_0 +i\Gamma/2)
 (\epsilon - \epsilon_0  - \Delta + \Delta_{L/R} +i\Gamma/2)}  }.
\label{eq:ALRpulse}
\end{equation}
Expression (\ref{eq:ALRpulse}) for $\ALR(\epsilon,t)$
applies for times, $t>s$, after the pulse has ended.
For times during the pulse, $0 < t < s$, one should replace $s$ by $t$,
which follows from causality.
In Fig. 1a, the current flowing
through the barriers is plotted for a voltage pulse of duration
$s = 3 \hbar/\Gamma$ (dashed curve).
Before the pulse, the chemical potentials $\mulr$ and
the level energy $\e_0$ are equal so the current is zero.
During the pulse, energies in the left lead are increased
by $\DeltaL = 10 \Gamma$ and the energy of the resonant level is raised
by $\Delta = 5 \Gamma$, appropriate for a symmetric structure
(inset, Fig. 1b).
(The choice $\DeltaR = 0$ is arbitrary since only the
relative shifts of energy are significant.)
Initially the current through each barrier grows proportional
to $t {\rm log}(1/t)$,$^{\tlogt}$ and then oscillates with a period
$\Delta t = 2\pi\hbar/|(\mu_{\ss L/R} +\DeltaLR) - (\epsilon_0 + \Delta)|
\ (\simeq 1.26$ in Fig. 1a). The time scale for the decay of
the oscillations is the resonance lifetime $\hbar/\Gamma$, {\it i. e.},
the time during which the states comprising the resonance remain
in phase and hence contribute constructively to the current
oscillations.
For the case shown in Fig. 1a, the occupancy remains fixed
at $n = 0.5$ by symmetry, and so the currents through the two
barriers must be equal. The ``ringing" in the current is
a consequence of the different phase evolution of
the resonant level and of the states in the leads.
Such phase coherence is
explicitly absent in the calculation of time-dependent
resonant-tunneling currents by Chen and Ting,$^{\chena}$
and so they find only exponential relaxation following
an abrupt change of bias.

To observe the ``ringing" experimentally,
a series of pulses such as that of
Fig. 1a could be applied to a tunneling structure
and the dc-current measured as a function of pulse
duration.$^{\leoa}$ In Fig. 1b, the derivative of the dc-current with
respect to pulse length is plotted, normalized by the repeat
time, $\tau$, between pulses.  For pulse lengths, $s$, of the
order of the resonance lifetime, $\hbar/\Gamma$, the derivative
of the dc-current mimics closely the time-dependent
current following the pulse, and, likewise, asymptotes to
the steady-state current at the new voltage.
We assume that the pulse duration,
of order a typical resonance lifetime, $\hbar/\Gamma \sim 1-1000 {\rm ps}$,
will be much shorter than the time between pulses, so that
pulses will not interfere.

Instead of applying a bias pulse, it may be experimentally
more practical to apply an ac-bias to the tunneling structure.$^{\dbrts}$
{}From Eq. (\ref{eq:ALR}), one finds for
an ac-potential,
\begin{equation}
A_{L/R}(\epsilon,t) = \displaystyle{\exp \Bigl[ -i{{\Delta - \DeltaLR}
              \over{\hbar\w}}
      \sin(\w t) \Bigr]}
  \displaystyle{\sum_{k= -\infty}^{\infty} J_k \Bigl({{\Delta - \DeltaLR}
              \over{\hbar\w}}\Bigr)
      { {\exp( i k \w t)}
              \over {\e - \e_0 -  k \hbar\w + i \Gamma/2} } },
\label{eq:ALRac}
\end{equation}
where
$\DeltaLR(t) = \DeltaLR \cos(\w t)$,
$\Delta(t) = \Delta \cos(\w t)$,
and $J_k$ is the $k^{th}$ order Bessel function.
In Fig. 2a, the current is plotted for an ac-potential
of frequency $\w=2\Gamma/\hbar$. While the current
has the same period, $2\pi/\w$, the complex time-dependence inside
each period is similar to the ``ringing" response to a pulse shown
in Fig. 1. One consequence of this complex harmonic structure is that
for temperatures $k_{\sss B}T < \hbar\w$
the time-averaged current, $\Jdc$, oscillates as a function of
period, as shown in Fig. 2b. An oscillation occurs whenever
a photon-assisted resonant-tunneling peak aligns with one chemical
potential, $\e_0 + k \hbar \w = \mulr$, giving oscillations
periodic in $1/\w$.$^{\sokol}$
Direct measurement of the ac-current
is complicated by the capacitive contributions from the leads,
but we expect the structure shown in Fig. 2a to survive.

In conclusion, a general formula (\ref{eq:JLb}) for the
time-dependent current through an interacting mesoscopic region
has been obtained in terms of local Green functions. While for
the interacting case only approximate Green function solutions
may be available, for a single non-interacting level with
energy-independent coupling to two leads an exact solution
has been obtained (\ref{eq:Grb} - \ref{eq:ALR}). This case
corresponds experimentally to resonant tunneling through a
double-barrier structure. We find that temporal
coherence of electrons tunneling through the resonant level
leads to ``ringing" of the current in response to an abrupt
change of bias, and similarly complex time-dependence in response to
an ac-bias.  This ``ringing" can be observed
experimentally in the dc-current by varying the pulse length in
a train of voltage pulses.
We hope that time-dependence will provide a new window on coherent
quantum transport and will lead to significant new insights in the future.

\acknowledgments
We thank S. J. Allen, M. Kastner, and L. Kouwenhoven
for valuable discussions. One of us (NSW) gratefully acknowledges
the NORDITA mesoscopic program for hospitality during early
stages of this work. Work at UCSB was supported by NSF
grant no. NSF-DMR90-01502
 and by the
NSF Science and Technology Center for Quantized Electronic Structures,
Grant no. DMR 91-20007.
\newpage
Figure Captions:\\
(1a) \, Time-dependent current, $J(t)$,
through a symmetric double-barrier tunneling
structure in response to a rectangular bias pulse.
All energies are in units
of the elastic coupling to the leads, $\Gamma$,
the current is in units of $e\Gamma/\hbar$, and all times are
in units of $\hbar/\Gamma$. Initially, the chemical potentials $\mul$ and
$\mur$ and the resonant-level energy $\e_0$ are all zero. At $t=0$,
a bias pulse (dashed curve)
suddenly increases energies in the left lead by $\DeltaL = 10$ and increases
the resonant-level energy by $\Delta = 5$ (see inset).
At $t=3$, before the current
has settled to a new steady value, the pulse ends and the current decays
back to zero. By symmetry, the
occupancy of the resonant level is always 0.5 and the currents through
the two barriers are equal.
The temperature in all figures is $k_{\sss B}T = 0.1 \Gamma$.
(1b) \, Derivative of the integrated
dc-current, $\Jdc$, with respect to pulse duration, $s$, normalized
by the interval between pulses, $\tau$. For pulse
durations much longer than the resonance lifetime $\hbar/\Gamma$,
the derivative is just
the steady-state current at the bias voltage, but for shorter pulses
the ``ringing" response of the current is evident.

(2a) Time-dependent current, $J(t)$, through a symmetric
double-barrier tunneling structure for an ac-bias
of frequency $\w = 2\Gamma/\hbar$ (dashed curve). The ac-driving amplitude is
$\DeltaL = 10$ about $\mul = 10$ in the left lead, $\Delta = 5$
about $\e_0 = 5$ for the level, and $\DeltaR = \mur = 0$ in the right
lead (see inset). By symmetry $n = 0.5$ and
the currents through the two barriers are equal.
(2b) \, Time-averaged current, $\Jdc$, as a function
of the ac-oscillation period $2\pi/\w$.
The~ac-amplitudes~are~the~same~as~those~used~in~(a).
\newpage
References
\par\noindent
$^*$Also at: MIC, Technical University of Denmark, DK-2800 Lyngby,
Denmark
\par\noindent
\begin{enumerate}
\item For a review see \quote{P. A. Lee and T. V. Ramakrishnan}
{Rev. Mod. Phys.}{57}{287 (1985)}
\item For a series of review articles see {\sl Mesoscopic
Phenomena in Solids}, B. L. Altshuler, P. A. Lee, and
R. A. Webb, eds. (Elsevier, Amsterdam, 1991).
\item \qquote{J. F. Whitaker {\it et al.}}
{\apl}{53}{385 (1988)} \quote{E. R. Brown {\it et al.}}
{\apl}{58}{2291 (1991)}
\item \qquote{L. P. Kouwenhoven {\it et al.}}
{\prl}{67}{1626 (1991)}
for a recent review see {\sl Single Charge Tunneling}, H. Grabert,
J. M. Martinis, and M. H. Devoret, eds. (Plenum, New York, 1991).
\item P. S. S. Guimaraes {\it et al.}, preprint.
\item A linear-response formula for the time-dependent non-interacting
current was recently derived by \quote{Y. Fu and S. C. Dudley}
{\prl}{70}{65 (1993)}
\item The time-averaged current in response to an ac-bias has
been analyzed previously for this model:
\qquote{D. Sokolovski}{\prb}{37}{4201 (1988)}
P. Johansson, {\sl Phys. Rev. B}{\bf 41}, 9892 (1990). Johansson also
considers the time-dependent current in a phenomenological
approach.
\item \quote{Y. Meir and N. S. Wingreen}{\prl}{68}{2512 (1992)}
\item \qquote{L. V. Keldysh}{Sov. Phys. JETP}{20}{1018 (1965)}
\quote {C. Caroli {\it et al.}}{J. Phys.}{C 4}{916 (1971)}
\item \book {G. D. Mahan} {Many-Particle Physics}{(Plenum Press, New
York, 1990)}
\item \quote{N. S. Wingreen, K. W. Jacobsen, and J. W. Wilkins}
{\prb}{40}{11834 (1989)}
\item \quote {K. L. Jensen and F. A. Buot}{\prl}{66}{1078 (1991)}
\item The initial change in current (\ref{eq:JLR}) is due to a change in
$\ALR(\epsilon,t)$. The change, $\delta \ALR(\epsilon,t)$,
has an overall amplitude proportional to $t$, and a long tail ( $\propto
1/\epsilon$) out to energies $|\epsilon| \sim \hbar/t$, which leads to a
growth of the current as
$\displaystyle{ t\!\!\int_{-\hbar/t}^\mu \!\!d\epsilon/\epsilon
\sim t {\rm log}(1/t)}$.
\item \qquote{L. Y. Chen and C. S. Ting}{\prl}{64}{3159 (1990)}
 {\sl \prb}{\bf 43}, 2097 (1991).
\item Leo Kouwenhoven, private communication; however, to observe
the ``ringing" experimentally requires a pulse rise-time faster than
the ringing period.
\end{enumerate}

\end{document}